\documentstyle[12pt]{article}

\begin{document}



\title{A modern review of the two-level approximation}

\author{Marco Frasca \\
Via Erasmo Gattamelata, 3, \\
         00176 Roma (Italy)}

\date{\today}

\maketitle

\abstract{
The paradigm of the two-level atom is revisited and its perturbative analysis is discussed
in view of the principle of duality in perturbation theory. The models we consider are
a two-level atom and an ensemble of two-level atoms both interacting with a single radiation mode. 
The aim is to see how the latter can be actually
used as an amplifier of quantum fluctuations to the classical level through the
thermodynamic limit of a very large ensemble of two-level atoms [M. Frasca, Phys. Lett.
A {\bf 283}, 271 (2001)] and how can remove Schr\"odinger cat states. 
The thermodynamic limit can be very effective for producing
both classical states and decoherence on a quantum system that evolves without dissipation.
Decoherence without dissipation is indeed an effect of a single two-level atom interacting
with an ensemble of two-level atoms, a situation that proves to be useful to understand
recent experiments on nanoscale devices showing unexpected disappearance of
quantum coherence at very low temperatures.
}



\newpage

\section{Introduction}

It is safe to say that the foundations of quantum optics are built on the
concept of a few level atom. Indeed, the most important concept introduced so
far in this field is the two-level atom \cite{ebe}. A lot of physics can be
derived by such an approximation and several recent experiments agree fairly
well with a description given by the so-called Jaynes-Cumming model describing
a two-level atom interacting with a single radiation mode \cite{exp}. Besides,
by this understanding of radiation-matter interaction it has also been possible
to generate Fock states of the radiation field on demand \cite{walt}.

The radiation-matter interaction currently used is based on some relevant
approximations that are still well verified in current experiments: Firstly, it
is assumed that the dipole approximation holds, that is, the wavelength of the radiation
field is much larger than the atomic dimensions; Secondly, the rotating
wave approximation (RWA) is always assumed, meaning by this that just near resonant terms
are effective in describing the interaction between radiation and matter,
these terms being also described as energy conserving. Indeed,
it is sometimes believed that, without these two approximations no two-level atom approximation
can really holds \cite{scu}. Actually, in the optical regime, that statement
can be supported and widely justifies the success of the Jaynes-Cumming
model both theoretically and experimentally.

Actually, things are not so straightforward to describe radiation-matter interaction.
Infact, Cohen-Tannoudji and coworkers were forced to introduce the concept
of dressed states for the two-level atom \cite{ct} as, in the regime of microwaves,
the RWA fails and a good description of the first
experiments in this field were achieved through the concept of dressed states
without the RWA \cite{ct}. 

Quantum computation exploited by ionic traps has been firstly put out by
Cirac and Zoller \cite{cz}. A recent paper by Moya-Cessa et al \cite{mc} proved
that the standard Jaynes-Cummings model should retain all terms for a Paul trap
giving a clear example of dismissal of the rotating wave approximation in
quantum optics.

The appearance of laser sources that have large intensity has made thinkable the possibility
to extend the study of a two-level atom in such a field. Recent studies seem to
indicate that such an approximation can give a viable model for such a physical
situation \cite{tlac1,tlac2,tlac3,tlac4,tlac5,tlac6,tlac7}. 
In view of this possibility, some methods have been
recently devised to approach a solution of the two-level atom in a monochromatic
field (being the laser field treated classically) \cite{tla1,tla2,tla3}. 
These studies retain just the two-level and dipole approximations but give up the RWA.

In our recent analysis, it was shown that, treating the laser field classically in
this situation, leaves out a relevant part of the behavior of the model \cite{fra1,fra2}.
Particularly, if one is especially interested in a resonant behavior, it is seen that some
Rabi oscillations are neglected: These oscillations have been recently observed in an
experiment with Josephson junctions \cite{naka} and originate from the 
formation of bands for the two levels of the atom due to the radiation
field \cite{fra2}. 

The aim of this paper is to review, using the approach of duality in
perturbation theory \cite{fra3}, the consequences of the validity of the two-level
approximation relaxing the RWA approximation. We will see that
a single radiation mode interacting with a large number of two-level atoms,
without the RWA, provide the amplification of the quantum fluctuations of
the ground state of the radiation mode producing a classical radiation field \cite{fra4}
and is able to remove macroscopic quantum superposition states.
It is important to point out that these effects arise when the 
initial state of the ensemble of two-level atoms is properly prepared and the
way of generating classical states by unitary evolution in the thermodynamic
limit of Ref.\cite{fra4} is considered. Non dissipative decoherence can also appear
as interaction between an ensemble of two-level systems
and a quantum system interacting with it \cite{fra7}.
It should be said that another approach to non dissipative decoherence 
has been recently proposed by Bonifacio and coworkers \cite{bo}.

The paper is so structured. In section \ref{sec1} we analyze the model from
a general perspective deriving the two-level approximation. In section \ref{sec2} 
we present a perturbative analysis of the two-level atom 
interacting with a single radiation mode, by
duality in perturbation theory. In section \ref{sec3} we give a brief survey of
a recent proposal of appearance of classical states and decoherence by unitary evolution
in the thermodynamic limit.
In section \ref{sec5} we show how a strong radiation field can be obtained by
strong interaction of a single mode with an ensemble of two-level atoms. 
In section \ref{sec5b} we present a way to approach the measurement problem in
quantum mechanics showing how, in the thermodynamic limit, Schr\"odinger cat
states can be removed leaving only a coherent state describing a classical field.
Finally, in section \ref{sec6} the conclusions are given.

\section{A paradigm in quantum optics: Two-level atom}
\label{sec1}

In this section we want to derive the two-level approximation on a general
footing. So, let us consider a system described by a Hamiltonian $H_0$ such
that we have a complete set of eigenstates $H_0|n\rangle=E_n|n\rangle$. We assume,
for the sake of simplicity, that the set is discrete. Then, we introduce a
time-independent perturbation $V$. By using the identity 
$I=\sum_n|n\rangle\langle n|$ we can write the Hamiltonian $H=H_0+V$ as
\begin{equation}
    H=\sum_n(E_n+\langle n|V|n\rangle)|n\rangle\langle n|
	+\sum_{m \neq n}|m\rangle\langle n|\langle m|V|n\rangle.
\end{equation}
This Hamiltonian can be rewritten by introducing the operators
\begin{eqnarray}
     \sigma_{nm}&=&|n\rangle\langle m| \nonumber \\
	 \sigma_{nm}^\dagger &=& |m\rangle\langle n| \nonumber \\
	 \sigma_{nm}^3 &=& \frac{1}{2}(|n\rangle\langle n|-|m\rangle\langle m|)
\end{eqnarray}
and we can build the algebra of the Pauli matrices, currently named su(2), as it
is straightforward to verify that
\begin{eqnarray}
   \left[\sigma_{nm},\sigma_{nm}^\dagger\right] &=& 2i\sigma_{nm}^3 \nonumber \\
   \left[\sigma_{nm}^3,\sigma_{nm}^\dagger\right] &=& i\sigma_{nm}^\dagger \nonumber \\
   \left[\sigma_{nm}^3,\sigma_{nm}\right] &=& -i\sigma_{nm}. 
\end{eqnarray}
This permits us to prove that our Hamiltonian can be
rewritten as the sum of two-level Hamiltonians. Infact, if we change to
the interaction picture by the unitary transformation (we use units $\hbar=c=1$)
\begin{equation}
    U_0(t)=\exp\left(-it\sum_n \tilde E_n|n\rangle\langle n|\right),
\end{equation}
being $\tilde E_n = E_n+\langle n|V|n\rangle$, 
and we rewrite the $V$ term of the Hamiltonian as
\begin{equation}
    V' = \sum_{m \neq n}|m\rangle\langle n|\langle m|V|n\rangle=
	\sum_{m > n}\left[\langle m|V|n\rangle\sigma_{nm}^\dagger+
	\langle n|V|m\rangle\sigma_{nm},
	\right]
\end{equation}
we get
\begin{equation}
    H_I = U_0^\dagger(t)V'U_0(t)=
	\sum_{m > n}\left[
	e^{-i(\tilde E_m -\tilde E_n)t}\langle m|V|n\rangle\sigma_{nm}^\dagger+
	e^{-i(\tilde E_n -\tilde E_m)t}\langle n|V|m\rangle\sigma_{nm}
	\right] 
\end{equation}
that proves our assertion: {\em The time evolution of a quantum system can be
described by a Hamiltonian being the sum of su(2) Hamiltonians}. The reason why
this problem is not generally solvable arises from the fact that, given two
su(2) parts $H_{I,i}$ and $H_{I,j}$ of this Hamiltonian, it can happen that
$[H_{I,i},H_{I,j}]\neq 0$ and then, the time evolution is not straightforward
to obtain analytically. Anyhow, the Hamiltonian $H_I$ can be used to realize
some approximate study of a quantum system. The simplest way to get
an approximate solution is, indeed, the two-level approximation. 

The two-level approximation can be easily justified by assuming
that only nearest levels of the unperturbed atom really counts in the time
evolution, that is, the more the separation between levels is large and
the less important is the contribution to the time evolution of the system. This
means that terms with the weakest time dependence in $H_I$ are the most important.
Mathematically, this means that we assume a solution by a perturbation series
and recognize principally the terms where a slower time dependence is
present.

We now consider the case of a single radiation mode interacting with a system
having Hamiltonian $H_0$. This means in turn that we can choose
\begin{equation}
    V=\omega a^\dagger a + e\left(\frac{\omega}{2V}\right)^\frac{1}{2}(a^\dagger + a)x
\end{equation} 
being $e$ the electron electric charge,
$a$ and $a^\dagger$ the ladder operators of the radiation mode with frequency
$\omega$ and normalization volume $V$, and $x$ the coordinate where the field 
is oriented, having chosen a linear polarization for it. The dipole approximation 
is taken to hold. In this case we have
\begin{equation}
    H_I=\omega a^\dagger a + e\left(\frac{\omega}{2V}\right)^\frac{1}{2}
	\sum_{m > n}\left[
	e^{-i(\tilde E_m -\tilde E_n)t}\langle m|x|n\rangle\sigma_{nm}^\dagger+
	e^{-i(\tilde E_n -\tilde E_m)t}\langle n|x|m\rangle\sigma_{nm}
	\right](a^\dagger + a) 
\end{equation}
and we are now in a position to obtain the Jaynes-Cummings model. Indeed, we
can apply a new unitary transformation to the interaction picture
$U_0'=\exp\left(-it\omega a^\dagger a\right)$ to get
\begin{equation}
    H_I'=e\left(\frac{\omega}{2V}\right)^\frac{1}{2}
	\sum_{m > n}\left[
	e^{-i(\tilde E_m -\tilde E_n)t}\langle m|x|n\rangle\sigma_{nm}^\dagger+
	e^{-i(\tilde E_n -\tilde E_m)t}\langle n|x|m\rangle\sigma_{nm}
	\right](e^{-i\omega t}a^\dagger + e^{i\omega t}a). \label{eq:H}
\end{equation}
Now, on the basis of the two-level approximation given above, we have to
conclude that the only terms to retain are those having no time
dependence at all, and these are the resonant terms. Here, we recover the
rotating wave approximation (RWA). So, if we have two resonant
levels $m=2>n=1$ and we choose the phases of the eigenstates of the unperturbed
system so that $\langle m|x|n\rangle$ is real, we can finally
write the Hamiltonian of the Jaynes-Cummings model as
\begin{equation}
    H_{JC}=g(\sigma_{12}a^\dagger+\sigma_{12}^\dagger a)
\end{equation}
being $g=\langle 2|x|1\rangle e\left(\frac{\omega}{2V}\right)^\frac{1}{2}$
and the resonance condition $E_2-E_1=\omega$. This gives a proper understanding of
the success of the two-level atom approximation in quantum optics when weak fields
are involved. It is important to note that also a small detuning can be kept, in
agreement with the above discussion.

The Jaynes-Cummings model is good until the other terms in the Hamiltonian are
truly negligible. The higher order corrections can be computed by a quite
general approach as shown in Ref.\cite{fra6}. These turn out to be
corrections to the the Hamiltonian at the resonance (e.g. Bloch-Siegert shift and/or
a.c. Stark shift) plus the need to add higher orders of the small perturbation theory
to the solution. In the optical regime it is all negligible.

So, as the small perturbation theory plays a crucial role
in this analysis, one may ask what one can say if the perturbation $V$ 
becomes strong. Again, by assuming that only a su(2) component really contributes
to the Hamiltonian (\ref{eq:H}) we need to treat the Hamiltonian
\begin{equation}
    H_S'=\omega a^\dagger a + g\left[
	    e^{-i(\tilde E_1 -\tilde E_2)t}\sigma^\dagger_{12}+
	    e^{-i(\tilde E_2 -\tilde E_1)t}\sigma_{12}
	    \right](a^\dagger + a).
\end{equation}
By undoing the interaction picture transformation, this Hamiltonian can be
rewritten as
\begin{equation}
    H_S=\omega a^\dagger a + \frac{\Delta}{2}\sigma_3+g\sigma_1(a^\dagger + a), \label{eq:HS}
\end{equation}
having set $\sigma_{12}+\sigma^\dagger_{12}=\sigma_1$, $\sigma_3=2\sigma^3_{12}$ and $\Delta=E_2-E_1$.
Neither small perturbation theory nor rotating wave approximation apply. Our
aim in the next section is to discuss the perturbative solution of the
Schr\"odinger equation with this Hamiltonian. But, while for the case of the
Jaynes-Cummings Hamiltonian we have a fully theoretical 
justification for our approximation, in the strong coupling regime, the two-level 
approximation can be satisfactorily justified only by experiment, unless it is exact.

\section{Perturbative analysis of an interacting two-level atom}
\label{sec2}

In this section we will give a brief overview of the perturbative solution for a
system described by the Hamiltonian (\ref{eq:HS}) in the strong coupling regime.
This approach has been described in Ref.\cite{fra2}. To agree about what a
strong coupling regime should be, one has properly to define the weak coupling
regime. Indeed, if one has the Hamiltonian
\begin{equation}
      H = H_0 + \tau V
\end{equation}
being $\tau$ an ordering parameter, the weak coupling regime is the one with
$\tau$ very small ($\tau\rightarrow 0$), while the strong coupling regime is the one with
$\tau$ very large ($\tau\rightarrow\infty$). The duality principle in perturbation theory as devised
in Ref.\cite{fra3} permits to do perturbation theory in both the cases, if one
is able to find the eigenstates of $V$, supposing known those of $H_0$. Indeed, small perturbation theory
by the usual Dyson series gives (we set $\tau=1$ as this parameter is arbitrary)
\begin{equation}
     |\psi(t)\rangle = U_0(t){\cal T}\exp\left[-i\int_0^tV_I(t)\right]
\end{equation}
being $\cal T$ the time-ordering operator,
\begin{equation}
     U_0(t)=\exp\left(-itH_0\right)
\end{equation}
the time evolution of the unperturbed Hamiltonian, and
\begin{equation}
     V_I(t)=U_0^\dagger(t)VU_0(t)
\end{equation}
the transformed perturbation. The choice of a perturbation and an unperturbed part
is absolutely arbitrary. So, we can exchange the role of $H_0$ and $V$,
obtaining the dual Dyson series
\begin{equation}
     |\psi(t)\rangle = U_F(t){\cal T}\exp\left[-i\int_0^tH_{0F}(t)\right]
\end{equation}
being
\begin{equation}
     U_F(t)=\exp\left(-itV\right)
\end{equation}
the time evolution of the unperturbed Hamiltonian, and
\begin{equation}
     H_{0F}(t)=U_F^\dagger(t)H_0U_F(t)
\end{equation}
the transformed perturbation. The duality principle states that, when this exchange
is done, restating $\tau$, the series one obtains have the ordering parameters $\tau$
and $\frac{1}{\tau}$ respectively. One is the inverse of the other. So, if
we have the eigenstates of $V$ as $|v_n\rangle$ and eigenvalues $v_n$, one
can write
\begin{equation}
    U_F(t)=\sum_n e^{-iv_n t} |v_n\rangle\langle v_n|.
\end{equation} 
If $V$ is time dependent one has formally to rewrite the above as the adiabatic
series introducing the geometric phases of the eigenvectors that now could be
time dependent themselves \cite{fra3}.

Coming back to the Hamiltonian (\ref{eq:HS}), we realize that small perturbation
theory can be recovered if the unperturbed part is that of the two-level atom,
otherwise one has a strong coupling perturbation series with an unperturbed
Hamiltonian given by
\begin{equation}
     V=\omega a^\dagger a + g\sigma_1(a^\dagger + a).
\end{equation}
The dressed states originating by diagonalizing this Hamiltonian are well known
\cite{ct} and are given by
\begin{equation}
    |v_{n,\lambda}\rangle = |\lambda\rangle e^{\frac{g}{\omega}\lambda(a-a^\dagger)}|n\rangle
\end{equation}
being $\sigma_1|\lambda\rangle = \lambda|\lambda\rangle$ with $\lambda = \pm 1$ and,
$|n\rangle$ the Fock number states that are displaced by the exponential operator
\cite{kn}. The eigenvalues are $E_n=n\omega-\frac{g^2}{\omega}$ and are degenerate
with respect to $\lambda$. So, one has
\begin{equation}
    U_F(t)=\sum_{n,\lambda}e^{-iE_nt}|v_{n,\lambda}\rangle\langle v_{n,\lambda}|
\end{equation}
and the transformed Hamiltonian becomes
\begin{equation}
    H_{0F}=U_F^\dagger (t)\frac{\Delta}{2}\sigma_3U_F(t)=H'_0+H_1.
\end{equation}
Using the relation \cite{kn}
\begin{equation}
    \langle l|e^{\frac{g}{\omega}\lambda(a-a^\dagger)}|n\rangle =
	\sqrt{\frac{n!}{l!}}\left(\lambda\frac{g}{\omega}\right)^{l-n}
	e^{-\lambda^2\frac{g^2}{2\omega^2}}
	L_n^{(l-n)}\left(\lambda^2\frac{g^2}{\omega^2}\right) \label{eq:kn}
\end{equation}
with $l\ge n$ and $L_n^{(l-n)}(x)$ the associated Laguerre polynomial, one gets
\begin{equation}
    H'_0=\frac{\Delta}{2}\sum_n e^{-\frac{2g^2}{\omega^2}}
	L_n\left(\frac{4g^2}{\omega^2}\right)
	\left[
	|[n;\alpha_1]\rangle\langle[n;\alpha_{-1}]||1\rangle\langle -1|+
	|[n;\alpha_{-1}]\rangle\langle[n;\alpha_1]||-1\rangle\langle 1|
	\right]
\end{equation}
being $L_n$ the n-th Laguerre polynomial, 
$|[n;\alpha_\lambda]\rangle=e^{\frac{g}{\omega}\lambda(a-a^\dagger)}|n\rangle$, and
\begin{eqnarray}
    H_1=\frac{\Delta}{2}\sum_{m,n,m\neq n}e^{-i(n-m)\omega t}
	\left[
	\langle n|e^{-\frac{2g}{\omega}(a-a^\dagger)}|m\rangle
	|[n;\alpha_1]\rangle\langle[m;\alpha_{-1}]||1\rangle\langle -1|+
	\right.\nonumber \\
	\left.
	\langle n|e^{\frac{2g}{\omega}(a-a^\dagger)}|m\rangle
	|[n;\alpha_{-1}]\rangle\langle[m;\alpha_1]||-1\rangle\langle 1|
	\right].
\end{eqnarray}
The Hamiltonian $H'_0$ can be straightforwardly diagonalized with the eigenstates
\begin{equation}
    |\psi_n;\sigma\rangle=\frac{1}{\sqrt{2}}
	\left[
	\sigma|[n;\alpha_1]\rangle|1\rangle+
	|[n;\alpha_{-1}]\rangle|-1\rangle
	\right]
\end{equation}
and eigenvalues
\begin{equation}
    E_{n,\sigma}=\sigma\frac{\Delta}{2}e^{-\frac{2g^2}{\omega^2}}
	L_n\left(\frac{4g^2}{\omega^2}\right)
\end{equation}
being $\sigma=\pm 1$. We see that two bands of levels are formed and two kind
of transitions are possible: interband (between levels of the two bands) and intraband (between
the levels of a band). This cannot happen if we consider a classical radiation mode, 
the intraband transitions would be neglected.
So, looking for a solution in the form
\begin{equation}
    |\psi_F(t)\rangle=\sum_{\sigma,n}e^{-iE_{n,\sigma}t}a_{n,\sigma}(t)|\psi_n;\sigma\rangle
\end{equation}
one gets the equations for the amplitudes \cite{fra2}
\begin{equation}
    i\dot{a}_{m,\sigma'}(t)=\frac{\Delta}{2}\sum_{n \neq m,\sigma}a_{n,\sigma}(t)
	e^{-i(E_{n,\sigma}-E_{m,\sigma'})t}e^{-i(m-n)\omega t}
	\left[\langle m|e^{-\frac{2g}{\omega}(a-a^\dagger)}|n\rangle\frac{\sigma'}{2}+
	\langle m|e^{\frac{2g}{\omega}(a-a^\dagger)}|n\rangle\frac{\sigma}{2}
	\right]. \label{eq:amp}
\end{equation}
This equations can also display Rabi oscillations between the eigenstates
$|\psi_n;\sigma\rangle$ that can be seen as macroscopic quantum superposition
states, both for interband and intraband transitions \cite{fra2}. 
States like these could prove useful for quantum computation. These kind of Rabi
oscillations in Josephson junctions have been recently observed \cite{naka}.

At this stage it is very easy to do perturbation theory in the strong coupling
regime for this model. One has to rewrite the initial condition $|\psi(0)\rangle$
by the eigenstates $|\psi_n;\sigma\rangle$ obtaining in this way 
the amplitudes $a_{n,\sigma}(0)$. Then, one has to solve eq.(\ref{eq:amp})
perturbatively as done routinely in the weak coupling regime. In case of a
resonance one has to apply the RWA obtaining Rabi oscillations. In this way we
see that the dual Dyson series, as it should be expected, displays all the
features of the standard weak coupling expansion.

\section{Classical states and decoherence by unitary evolution}
\label{sec3}

An ensemble of independent two-level systems can behave classically. This has
been proven in Ref.\cite{fra4}. Indeed, let us consider a Hamiltonian
\begin{equation}
    H_c = \frac{\Delta}{2}\sum_{i=1}^N \sigma_{3i}.
\end{equation}
Assuming distinguishable systems, we can take for the initial state the one given by
\begin{equation} 
|\psi(0)\rangle=\prod_{i=1}^N(\alpha_i|\downarrow\rangle_i+\beta_i|\uparrow\rangle_i) \label{eq:is}
\end{equation} 
with $|\alpha_i|^2+|\beta_i|^2=1$, $\sigma_{z i}|\downarrow\rangle_i=-|\downarrow \rangle_i$
and $\sigma_{z i}|\uparrow\rangle_i=|\uparrow \rangle_i$. The time evolution gives us
\begin{equation} 
|\psi(t)\rangle=\prod_{i=1}^N(\alpha_i e^{i \frac{\Delta}{2}t}|\downarrow \rangle_i+
\beta_ie^{-i\frac{\Delta}{2}t}|\uparrow\rangle_i).
\end{equation}
For the Hamiltonian it is easy to verify that
\begin{equation}
    \langle H_c \rangle = \langle\psi(0)|H_c|\psi(0)\rangle = 
	\frac{\Delta}{2}\sum_{i=1}^N (|\beta_i|^2-|\alpha_i|^2)=
	\frac{\Delta}{2}k_H N
\end{equation} 
being $k_H$ a fixed number between $-1$ and $1$. So, in a similar way, it easy
to obtain the fluctuation
\begin{equation}
(\Delta H_c)^2=
\langle\psi(0)|H_c^2|\psi(0)\rangle-\langle\psi(0)|H_c|\psi(0)\rangle^2=\Delta^2k'_H N.
\end{equation}
being $k'_H=\sum_{i=1}^N|\beta_i|^2(1-|\beta_i|^2)/N$
and one sees that $k'_H$ is a finite number independent on $N$.
So, as it happens in statistical thermodynamics, in the thermodynamic limit $N\rightarrow\infty$
we see that quantum fluctuations are not essential, that is
\begin{equation}
    \frac{\Delta H_c}{H_c} \propto \frac{1}{\sqrt{N}}.
\end{equation}
The ``laws of thermodynamics'' are obtained by the Ehrenfest's theorem
and are the classical equations of motion. That is,
the variables $\Sigma_x=\sum_{i=1}^N \sigma_{x i}$, 
$\Sigma_y=\sum_{i=1}^N \sigma_{y i}$ and $\Sigma_z=\sum_{i=1}^N \sigma_{z i}$
follow, without any significant deviation, the classical equations of motion,
when the thermodynamic limit is considered and the time evolution is
computed averaging with the above $|\psi(t)\rangle$. So, we
have found a classical object out of the quantum unitary evolution. The main point
here is that {\sl classical objects can be obtained by unitary evolution in the
thermodynamic limit depending on their initial states}. Actually, one cannot apply
the above argument if e.g. the state of the system is an eigenstate of the Hamiltonian
$H_c$. Besides, a classical state obtained by unitary evolution, {\sl per se},
does not produce decoherence. Rather, it is interesting to see what happens when
such a classical state interacts with some quantum system. This is a relevant
problem that can prove quantum mechanics and its fluctuations to be just
the bootstrap of a classical world: If by unitary evolution,
in the thermodynamic limit, some classical objects
are obtained and these are permitted to interact with other quantum objects, the
latter can decohere or become classical by themselves.

As a relevant example, let us consider the interaction of the above system
with a two-level atom. This model has been considered in Ref.\cite{fra7} as a 
possible explanation of recent findings in some experiments
with nanoscale devices that show unexpected decoherence in the low temperature
limit \cite{T0,fe}. The Hamiltonian can be written as
\begin{equation}
    H_D=\frac{\Omega_0}{2}\sigma_z+
	\frac{1}{2}\sum_{i=1}^N (\Delta_{xi}\sigma_{xi}+\Delta_{zi}\sigma_{zi})
	-J\sigma_x\cdot\sum_{i=1}^N \sigma_{xi}. \label{eq:model}
\end{equation}
where $J$ is the coupling. The Hamiltonian of the two-level systems 
(second term in eq.(\ref{eq:model}))
is taken not diagonalized, but this does not change our argument
as the above analysis still applies. 
Finally, $\Omega_0$ is the parameter of the Hamiltonian of the two-level atom
that we want to study. We need another hypothesis to go on, that is, we assume
that the coupling $J$ is larger than any parameter of the two-level
systems $\Delta_{xi},\Delta_{zi}$, but not with respect to $\Omega_0$. 
By applying duality in perturbation theory, we have the leading order solution
\begin{equation}
    |\psi(t)\rangle \approx 
	\exp\left(\frac{-it\Omega_0}{2}\sigma_z+iJt\sigma_x\cdot\sum_{i=1}^N \sigma_{xi}\right)
	|\psi(0)\rangle.
\end{equation}
Now, as already seen, we have to choose the state of the two-level systems
as given by the product of the lower eigenstates of each $\sigma_{xi}$. This can be
seen as a kind of ``ferromagnetic'' state and is in agreement with our preceding discussion.
So, we take
\begin{equation}
    |\psi(0)\rangle = |\downarrow\rangle\prod_{i=1}^N |-1\rangle_i
\end{equation}
being $\sigma_z|\downarrow\rangle=-|\downarrow\rangle$ and, similarly,
$\sigma_z|\uparrow\rangle=|\uparrow\rangle$. The state of the ensemble of two-level
systems agrees fairly well with the one of eq.(\ref{eq:is}). So, one has, by
tracing away the state of the ensemble of two-level systems being not
essential for our aims,
\begin{equation}
    |\psi'(t)\rangle \approx 
	\exp\left(\frac{-it\Omega_0}{2}\sigma_z+iJNt\sigma_x\right)
	|\downarrow\rangle
\end{equation}
that defines a spin coherent state \cite{css,gil}. The point we are interested
in is the thermodynamic limit. When $N$ is taken to be large enough, the 
contribution $\Omega_0$ can be neglected and we have a reduced density matrix
\begin{equation}
    \rho'(t)=\exp\left(iJNt\sigma_x\right)|\downarrow\rangle
	\langle\downarrow|
	\exp\left(-iJNt\sigma_x\right)
\end{equation}
being
\begin{eqnarray}
    \rho'_{\uparrow\uparrow}(t)&=&\frac{1-\cos(2NJt)}{2} \\
	\rho'_{\uparrow\downarrow}(t) &=& -i\frac{1}{2}\sin{2NJt} \\
	\rho'_{\downarrow\uparrow}(t) &=& i\frac{1}{2}\sin{2NJt} \\
	\rho'_{\downarrow\downarrow}(t)&=&\frac{1+\cos(2NJt)}{2},
\end{eqnarray}
where we have oscillating terms with a frequency $NJ$ that goes to infinity
in the thermodynamic limit. The only meaning one can attach to such a frequency is by 
an average in time (see \cite{fra7} and Refs. therein) and decoherence is
recovered. So, when the ensemble of two-level systems strongly interacts with a quantum
system produces decoherence and quantum behavior disappears,
in the thermodynamic limit. The ensemble
of two-level systems should evolve unitarily, producing a classical behavior. Higher
order corrections have also been studied in Ref.\cite{fra7}. It is important
to stress that this behavior should be expected at zero temperature as quantum
coherence is lost otherwise.

Such a behavior, having a characteristic decoherence time scale depending on the number
of two-level systems that interact with the quantum one, has been recently
observed in quantum dots \cite{fe}. In this case, the ensemble of two-level
systems can be given by the spins of the electrons that are contained in the two dimensional electron gas
in the dot. Another source of decoherence in quantum dots
could be given by the spins of the nuclei interacting through an hyperfine interaction
with the spin of the conduction electrons \cite{loss}. The nuclei are contained in the
heterostructures forming the dot. In this case we have a similar spin-spin interaction but
isotropic. The mechanism that produces the decay of the
off-diagonal parts of the density matrix, also in this case, appears to be the same,
being the decoherence produced dynamically and dependent on the initial state.

\section{Amplification of quantum fluctuations to the classical level}
\label{sec5}

Spontaneous emission can be seen as a very simple example of decoherence in the
``thermodynamic limit'' of the number of radiation modes. Indeed, we can consider
a two-level system interacting with $N$ radiation modes and being resonant with
one of it. In the limit of a small coupling between radiation and two-level system
and very few spectator modes, one has Rabi oscillations, a clear example of
quantum coherence. When the number of spectator modes is taken to go to infinity,
a description with continuum is possible and this gives rise to decay, i.e.
spontaneous emission. This representation of the process of decay is very well
described in Ref.\cite{ct}.

Here, we want to consider the opposite situation, that is, a single radiation
mode strongly interacting with an ensemble of $N$ two-level systems. We are going
to show that, when the ensemble of two-level systems behaves as a classical
object if left alone, the radiation field, supposed initially in the ground
state, will have the zero point fluctuations amplified to produce a classical
field having intensity dependent on $N$, the number of two-level systems.

As done in Ref.\cite{fra4}, we modify the model of eq.(\ref{eq:HS}) to consider
$N$ two-level systems interacting with a single radiation mode, as
\begin{equation}
    H_S=\omega a^\dagger a + 
	\frac{\Delta}{2}\sum_{i=1}^N\sigma_{3i} +
	g\sum_{i=1}^N\sigma_{1i}(a^\dagger + a). \label{eq:HSi}
\end{equation} 
Then, the strong coupling regime amounts to consider the Hamiltonian
$\frac{\Delta}{2}\sum_{i=1}^N\sigma_{3i}$ as a perturbation, as already done
in sec.\ref{sec2} for a single two-level atom. We take as initial state
of the full system $|\psi(0)\rangle=|0\rangle\prod_{i=1}^N|-1\rangle_i$, so
that, the ensemble of two-level systems is again in a kind of ``ferromagnetic'' state
representing its ground state. Besides, no photon is initially present.
It is a well known matter that the fluctuations
of the radiation mode are not zero in this case. The unitary evolution at the
leading order gives us
\begin{equation}
    |\psi(t)\rangle \approx 
	\exp\left[-it\omega a^\dagger a-itg\sum_{i=1}^N\sigma_{1i}(a^\dagger + a)\right]
	|0\rangle\prod_{i=1}^N|-1\rangle_i
\end{equation}
that, by use of a known disentangling formula \cite{gil}, produces
\begin{equation}
    |\psi(t)\rangle = e^{i\xi(t)}e^{-i\omega a^\dagger at}D[\alpha(t)]|\psi(0)\rangle,
\end{equation}
being
\begin{equation}
    \xi(t)=\frac{N^2g^2}{\omega^2}(\omega t- \sin(\omega t)),
\end{equation}
\begin{equation}
    \alpha(t)=-\frac{Ng}{\omega}(1-e^{i\omega t}),
\end{equation}
and
\begin{equation}
    D[\alpha(t)]=\exp[\alpha(t)a^\dagger - \alpha(t)^* a].
\end{equation}
We conclude that, at the leading order, the radiation mode evolves
as a coherent state with a parameter given by
\begin{equation}
    \hat{\alpha}(t)=-\frac{Ng}{\omega}(e^{-i\omega t}-1)=\alpha(t)e^{-i\omega t}.
\end{equation}
In this way, we have amplified the quantum fluctuations of the field, being
the fluctuation of the number of photons proportional to $N$, but, as
the average of the number of photons is proportional to $N^2$, this ratio
goes to zero as the thermodynamic limit $N\rightarrow\infty$ is taken. 
As it is well know \cite{mw}, this
produces a classical field with increasing intensity as the number of two-level
systems increases, proving our initial assertion. We can see that the amplification
of quantum fluctuations gives rise to a classical object, as initially no
radiation field is present.

Higher order corrections have been studied in Ref.\cite{fra8}, showing that are
not essential in the thermodynamic limit. So, this effect will prove to be a
genuine example of production of a classical object by unitary evolution 
in the thermodynamic limit with possible technological applications. 

\section{Two-level Systems, Thermodynamic Limit and Schr\"odinger Cat States}
\label{sec5b}

Decoherence, as currently devised, is able to remove superposition states through
interaction of the environment with a quantum system. This does not solve the
measurement problem in quantum mechanics as, mixed forms of the density matrix
do not give single states required by the measurement process \cite{cri}. This
problem is fairly well described by the Schr\"odinger cat paradox as we ask that
the cat has a well defined state at the observation. Schr\"odinger cat states
have been currently produced in laboratory in a form of superposition of coherent states (see e.g. the 
the second reference in \cite{ebe})
\begin{equation}
    |\psi_{cat}\rangle = {\cal N}(|\alpha e^{i\phi}\rangle + |\alpha e^{-i\phi}\rangle) \label{eq:cat}
\end{equation}
being ${\cal N}$ a normalization factor, $\alpha$ and $\phi$ real numbers.
To understand the measurement problem, we would like to get a single state
out of such a superposition after unitary evolution, if possible. In this way,
we can show, at least in this case, that quantum mechanics is, indeed, a self-contained
theory. This possibility can be exploited by an ensemble of two-level systems interacting
with a single radiation mode in the thermodynamic limit.

In order to accomplish our aim, we consider again the Hamiltonian (\ref{eq:HSi})
with the initial condition (\ref{eq:cat}) multiplied by the ``ferromagnetic
state'', $|\phi\rangle=\prod_{i=1}^N|-1\rangle_i$, for the ensemble of atoms as
to have $|\psi(0)\rangle = |\psi_{cat}\rangle|\phi\rangle$. The unitary evolution,
assuming the Hamiltonian of two-level atoms as a perturbation, gives in this case
\cite{frap}
\begin{equation}
     |\psi(t)\rangle\approx e^{i\xi(t)}
	 {\cal N}(e^{i\phi_1(t)}|\beta(t)e^{-i\omega t} 
	 + \alpha e^{i\phi-i\omega t}\rangle 
	 + e^{i\phi_2(t)}|\beta(t)e^{-i\omega t} + \alpha e^{-i\phi-i\omega t}\rangle)|\phi\rangle. \label{eq:st}
\end{equation}
where use has been made of the property of the displacement operator for coherent states so to yield
\begin{equation}
    \xi(t)=\frac{N^2g^2}{\omega^2}(\omega t - \sin(\omega t)),
\end{equation}
\begin{equation}
    \beta(t) = \frac{Ng}{\omega}(1-e^{i\omega t})
\end{equation}
and
\begin{equation}
    \phi_1(t) = -i\frac{\alpha}{2}[\beta(t)e^{-i\phi}-\beta^*(t)e^{i\phi}],
\end{equation}
\begin{equation}
    \phi_2(t) = -i\frac{\alpha}{2}[\beta(t)e^{i\phi}-\beta^*(t)e^{-i\phi}],
\end{equation}
being the phases $\phi_1(t)$ and $\phi_2(t)$ generated by multiplication of two displacement operators.
In the thermodynamic limit $N\rightarrow\infty$ one gets the
macroscopic state $|\beta(t)e^{-i\omega t}\rangle$ and the cat state seems gone away.

Actually, we have a couple of problems before one can claim that, effectively, the cat
state has been removed. Firstly, all we have done is a displacement to infinity and no
decoherence seems to be implied in such an operation, so all the properties of a superposition
state have to be there anyway. Secondly, we have done perturbation theory and one has to
prove that, in the thermodynamic limit, higher order corrections are negligible.

The first question is answered immediately by computing the interference term in the
Wigner function of the state (\ref{eq:st}). In the thermodynamic limit such a term 
should become negligible. One has
\begin{eqnarray}
    W_{INT} &=&\frac{2}{\pi}
	\exp\left[-\left(x + \frac{\sqrt{2}Ng}{\omega}(1-\cos(\omega t)) - \sqrt{2}\alpha\cos(\phi)\cos(\omega t)\right)^2\right] \\ \nonumber
	&\times&\exp\left[-\left(p + \frac{\sqrt{2}Ng}{\omega}\sin(\omega t) + \sqrt{2}\alpha\cos(\phi)\sin(\omega t)\right)^2\right] \\ \nonumber
	&\times&\cos\left[2\sqrt{2}\alpha\sin(\phi)\left(p\sin(\omega t) - x\cos(\omega t)\right)
	+\alpha^2\sin(2\phi) + 
	8\alpha\frac{Ng}{\omega}\sin(\phi)(1-\cos(\omega t))\right].
\end{eqnarray}
This term has a quite interesting form as displays a term that rapidly oscillates in time
for $N$ becoming increasingly large. If such oscillations become too rapid, we can invoke
blurring in time to have these terms averaged away. Otherwise, as the Wigner function can
be measured, we will get a way to probe, immediately and by very simple means, Planck time
physics. So, we can safely claim that we have true decoherence in the thermodynamic limit
and the cat state is effectively removed generating a macroscopic classical state. It should
be said that ordinary decoherence is generally invoked for blurring in space \cite{berry} and
there is no reason to say that also blurring in time cannot occur. We can recognize here
the same argument used in order to obtain decoherence for the model of sec.\ref{sec3}
to explain decoherence in quantum dots. A sound mathematical basis for such an approach
is given in \cite{hardy}.

The second question can be straightforward answered by computing higher order corrections
through the strong coupling expansion discussed in sec.\ref{sec2}. The proof is
successfully accomplished in Ref.\cite{frap} and we do not repeat it here.

It is interesting to note that all the new points introduced so far for analyzing two-level
systems can conspire to generate a new view of the way measurement is realized in quantum
mechanics, possibly making the theory self-contained.

\section{Discussion and Conclusions}
\label{sec6}

We have presented a brief review about some new views on two-level systems. These
appear to be even more important today with a lot of new effects to be described
and experimentally observed. Paramount importance is acquiring the decoherence
due to an ensemble of two-level systems as it is becoming ubiquitous to different
fields of application as quantum computation and nanotechnology, fields that
maybe could merge. To face these new ways to see the two-level approximation, we
have exposed new mathematical approaches to analyze models in the strong
coupling regime. This regime has been pioneered by Bender and coworkers \cite{bend}
in quantum field theory in the eighties, but, with our proposal of duality in
perturbation theory, a possible spreading of such ideas to other fields is now
become possible. Indeed, a lot of useful results, as those presented here, are
obtained by this new approach and, hopefully, the future should deserve some
other interesting results. 

The idea of a non dissipative decoherence is also relevant due to the recent
findings in the field of nanodevices, where unexpected lost of quantum coherence
has appeared in experiments performed at very low temperatures. In these cases,
it appears as the standard idea of decoherence, meant as interaction of a quantum system 
with an external environment, 
seems at odds with some experimental results, even if 
an interesting proposal through the use of quantum fluctuations has been put
forward by B\"uttiker and coworkers \cite{butt}.

The conclusion to be drawn is that, today, a lot of exciting work at the
foundations of quantum mechanics is expecting us, giving insight toward new
understandings and methods and, not less important, applications.

\label{end}

\end{document}